\documentclass[aps,prb,twocolumn,showpacs,superscriptaddress,groupedaddress]{revtex4-1}
\usepackage[dvips]{graphics,graphicx}
\usepackage{epstopdf}
\usepackage{bbold}
\usepackage[usenames, dvipsnames]{color}
\usepackage{lipsum}
\usepackage{amssymb}
\usepackage{flushend}

\begin{document}
\title{Avoided crossings and bound states in the continuum in low-contrast dielectric gratings}
\author{Evgeny N. Bulgakov$^{1,2}$}
\author{Dmitrii N. Maksimov$^{1,2}$}
\affiliation{$^1$Reshetnev Siberian State University of Science and Technology, 660037, Krasnoyarsk,
Russia\\
$^2$Kirensky Institute of Physics, Federal Research Center KSC SB
RAS, 660036, Krasnoyarsk, Russia}
\date{\today}
\begin{abstract}
We consider bound states in the continuum (BICs) in low-contrast dielectric gratings. It is demonstrated
that the BICs originate from the reduced guided modes
on the effective dielectric slab with the permittivity equal
to the average permittivity of the dielectric grating. In case of isolated resonances the positions of BICs can be
found from two-wave dispersion relationships for guided leaky modes. In the case of the degeneracy between the two families of leaky modes the
system exhibits an avoided crossing of resonances. In the spectral vicinity of the avoided crossing the transmittance as well
as the emergence of BICs
is described in the framework of the generic formalism by Volya and Zelevinsky  [Physical Review C 67, 054322
(2003)] with a single fitting
parameter.
\end{abstract}
 \maketitle

\section{Introduction}

Controlling the localization of electromagnetic waves through engineering high quality resonances is of fundamental
importance
in electromagnetesm \cite{John12,*Marpaung13,*Qiao18,Hsu16}. Several platforms have been investigated from microwaves to optics towards minimizing
radiation losses including
whispering gallery modes \cite{Gorodetsky99,*Zhu09}, metasurfaces \cite{Al-Naib09,*Jansen11,*Singh14,*Krasnok18},
photonic crystal microcavities \cite{Tanabe07,*Noda07}
dielectric resonators \cite{Lepetit10,*Lepetit14,Rybin17a}, and Fabry-Perot structures \cite{Velha07,*Zain08}.
Among those platforms dielectric gratings (DG) have become an important instrument in optics
with various application relaying on high quality resonances \cite{Qiao18,*Chang-Hasnain12}. In particular
ultra-high quality resonances were demonstrated in the spectral vicinity of the avoided crossings of the
DG modes \cite{Karagodsky11}.

The utmost case of light localization in dielectric structures is the emergence of bound states in the
continuum (BICs), i.e. localized eigenmodes of Maxwell's equations with infinite quality factor
embedded into the continuous spectrum of the scattering
states \cite{Hsu16}. In the recent past the optical BICs were experimentally observed in all-dielectric
set-ups with periodically varying permittivity \cite{Plotnik,*Weimann13,*Hsu13,*Vicencio15,*Sadrieva17,*Xiao17,Foley14a}.
The BICs have been extensively studied in various types of grated structures ranging from the simplest case of
an array of rectangular bars in air \cite{Foley14a,Zhen2014,*Blanchard16,*Wang16b,*Ni16,*Taghizadeh17}
to substarate gratings \cite{Yoon15,*Cui16a,*Monticone17,*Wang16,*Bulgakov18} and grated fibers \cite{Gao17,Bulgakov17b}.

\begin{figure}[t]
\begin{center}
\includegraphics[width=.49\textwidth, height=.245\textwidth,trim={5.6cm 11.5cm 1.0cm 11.0cm},clip]{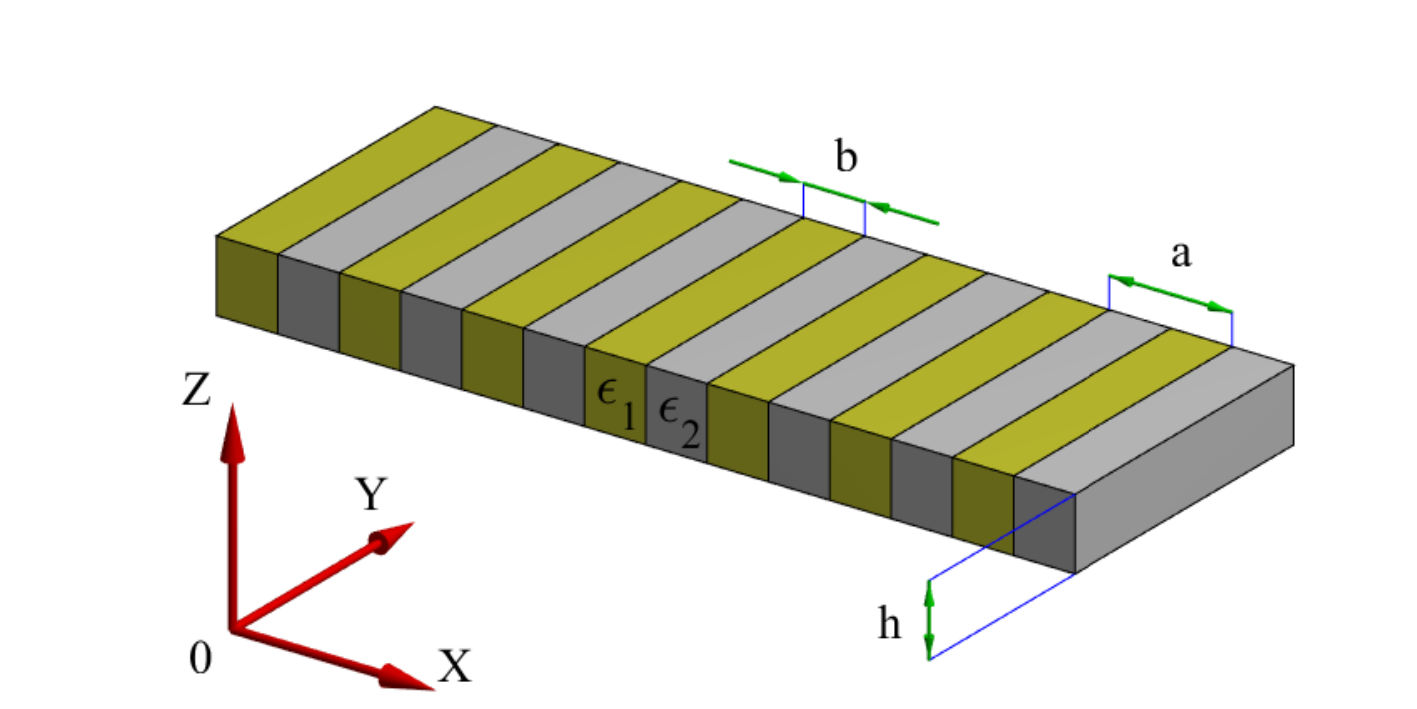}
\caption{(Color online) Grating composed of dielectric bars with different permittivities, $\epsilon_1$ and $\epsilon_2$ .}\label{Fig1}
\end{center}
\end{figure}

In this paper we consider BICs in planar DGs consisting of laterally arranged rectangular bars made of two
dielectric materials with a small difference in dielectric permittivity, see Fig. \ref{Fig1}.  Thus, the DG is a dielectric slab of
thickness $h$ in the $z$-direction with step-wise alternating permittivity with period $a$ along the $x$-axis.
 The bars with permittivity $\epsilon_1$
have thickness $b$ in the $x$-direction, while the bars with $\epsilon_2$ have thickness $a-b$. The grating is
infinitely extended in both $x$-, and $y$-directions.
Since the difference in permittivity between the neighboring bars is small with respect to the absolute values we shall
term such system {\it low contrast} DGs to emphasize the low dielectric contrast between the building blocks of the grating.
To avoid disambiguity we stress at the outset that the dielectric contrast between the grating itself and the surrounding
medium (air) can be arbitrary.

The generic mechanism of BIC formation was put forward in 1985 by Friedrich and Wintgen, who demonstrated \cite{Friedrich85}
that a BIC occurs as a product of destructive interference of two resonant modes coupled to the outgoing channel.
Here we show that the low-contrast DGs support BICs at any small dielectric contrast
between the bars due to interference between two non-orthogonal leaky modes \cite{Volya03} in the spectral vicinity of avoided
crossings. The results are verified against straightforward simulations with rigorous coupled wave
analysis (RCWA) \cite{Moharam81,*Tikhodeev02}.

Due to the system's translational symmetries the spectral parameters of the free-space eigenmodes are linked
through the following dispersion relationship \cite{Popov12}
\begin{equation}\label{spectrum}
k_0^2=k_{x,n}^2+k_z^2+k_y^2, \ k_{x,n}=\beta+2\pi{n}/a,
\end{equation}
where $k_0$ is the vacuum wave number, $k_{x,z}$ are the wave numbers along the $x,y$-axes,
$k_z$ is the far-field wave number in the direction orthogonal to the plane of the structure,
$\beta$ is the Bloch wave number (propagation constant), and, finally, $\ n=0,\pm 1, \ldots$ corresponds to the diffraction
order. Here, we consider transverse magnetic (TM) modes with $k_y=0$, i.e. propagating only perpendicular to the bars.
The analysis shall be performed in the spectral range
\begin{equation}\label{range}
\beta{a} < k_0{a} < 2\pi -\beta{a},
\end{equation}
which according to Eq. (\ref{spectrum}) means that only one TM scattering channel
is open in the far-zone on both sides of the DG.

The TM modes are the solutions of the scalar wave equation
\begin{equation}\label{wave}
\nabla^2\psi(x,z)+\epsilon(x,z)k_0^2\psi(x,z)=0,
\end{equation}
where $\epsilon(x,z)$ is the dielectric permittivity, and
$\psi(x,z)$ is the $y$-component of the electric vector, $\psi(x,z)=E_y(x,z)$. Following \cite{Gao17} we expand
 the dielectric permittivity into Fourier harmonics
\begin{equation}\label{eps_Fourier}
\epsilon(x,z)=\sum_{n=-\infty}^{\infty}\gamma_n e^{i2\pi nx/a}.
\end{equation}
In low-contrast DGs the following condition always holds true
\begin{equation}
|\gamma_0|\gg|\gamma_n|, n\neq 0.
\end{equation}
Retaining only $\gamma_0$ in Eq. (\ref{eps_Fourier})
would lead us to a uniform dielectric slab which does not support BICs.
Thus, for finding BICs both $\gamma_{\pm 1}$ should be retained in the first order approximation. We shall see in the next section, though,
that in {\it some} cases retaining only $\gamma_1$ (or $\gamma_{-1}$) suffices for finding the BICs
to a good accuracy.

\section{Two-wave BICs}

The solution for the EM field within the DG is written
\begin{eqnarray}\label{u_def}
\psi_{s}=U_{s}(x)\frac{\cos(\varkappa z)}{\cos(\varkappa  h/2)}e^{i\beta x}, \nonumber \\
\psi_{a}=U_{a}(x)\frac{\sin(\varkappa z)}{{\sin(\varkappa h/2)}}e^{i\beta x},
\end{eqnarray}
where $\varkappa$ is the propagation constant in the $z$-direction within the DG, and subscript $s,(a)$ stands for waves symmetric
 (antisymmetric) with respect to the central plane of the grating.
The functions
$U_{s,a}(x)$ obey the equation
\begin{equation}\label{u_eqn}
\left[\left(\frac{\partial}{\partial x} + i\beta\right)^2 + k_0^2\epsilon(x)\right]U_{s,a}(x)=\varkappa^2U_{s,a}(x).
\end{equation}
Let us write $U_{s,a}(x)$ in the following form
\begin{equation}\label{two_wave}
U_{s,a}(x)=\sum_{n=-\infty}^{\infty}u_{n}e^{i2\pi n x/a}.
\end{equation}
As shown \cite{Gao17} under condition that $\gamma_0$ and $\gamma_1$ are retained in Eq. (\ref{eps_Fourier}) we have to retain $u_0$, and $u_{-1}$ in
Eq. (\ref{two_wave}) to derive self-consistent equations from Eq. (\ref{u_eqn}). Those equations read \cite{Gao17}
\begin{eqnarray}\label{two_wave_eqn}
(\gamma_0 k_0^2-\beta^2)u_0+\gamma_1 k_0^2u_{-1}=\varkappa^2 u_0, \nonumber \\
\left[\gamma_0 k_0^2- \left( \beta-\frac{2\pi}{a} \right)^2 \right]    u_{-1}+\gamma_1 k_0^2u_0 =\varkappa^2 u_{-1}.
\end{eqnarray}

The symmetric (antisymmetric) waves can be excited if the the DG is symmetrically (antisymmetrically) illuminated
by monochromatic plane waves from the both sides. The general symmetric solution outside the grating $z>|h/2|$ reads
\begin{equation}\label{outside}
\psi(x,z)=e^{i\beta x}\left(A\frac{e^{-ik_{z,0}|z|}}{e^{-ik_{z,0}h/2}}
+\sum_n t_n \frac{e^{ik_{z,n}|z|+i2\pi nx/a}}{e^{ik_{z,n}h/2}} \right),
\end{equation}
where $A$ is the amplitude of the incident wave,
\begin{equation}
k_{z,n}=\sqrt{k_0^2-\left(\beta+\frac{2\pi n}{a} \right)^2},
\end{equation}
and $t_n$ are the amplitudes of the outgoing waves.
The antisymmetric solution only differs from Eq. (\ref{outside}) by its sign in the upper half space. To be consistent
with the two-wave approximation the summation runs $n=0,-1$. Thus, the total solution is given by four unknown
quantities $u_0, u_{-1}, t_0, t_{-1}$. By using  Eqs. (\ref{u_def}, \ref{two_wave}, \ref{two_wave_eqn}, \ref{outside})
together with the interface boundary conditions \cite{Gao17} we find

\begin{widetext}
\begin{equation}\label{big}
\left(
\begin{array}{cc}
J_0+J_{-1}\sigma_{-1}^2+(1+\sigma_{-1}^2)ik_{z,0} & \sigma_{-1}(J_{-1}-J_0) \\
\sigma_{-1}(J_{-1}-J_0) & J_{-1}+J_{0}\sigma_{-1}^2+(1+\sigma_{-1}^2)ik_{z,-1}
\end{array}
\right)
\left(
\begin{array}{c}
t_0 \\
t_{-1}
\end{array}
\right)=A
\left(
\begin{array}{c}
ik_{z,0}(1+\sigma_{-1}^2)-J_0-J_{-1}\sigma_{-1}^2 \\
-\sigma_{-1}(J_{-1}-J_0)
\end{array}
\right),
\end{equation}
\end{widetext}
where
\begin{equation}
J_n=\left\{
\begin{array}{l}
\varkappa_n \tan\left(\varkappa_n h/2 \right), \ {\rm symmetric\ waves}\\
-\varkappa_n \cot\left(\varkappa_n h/2 \right), \ {\rm antisymmetric\ waves}
\end{array}
\right.
\end{equation}
with
\begin{eqnarray}
\sigma_{-1}=\frac{\gamma_1 k_{0}^2}{f_{-1}^2-f_0^2}, \label{sigma} \\
\varkappa_{-1}^2=f_{-1}^2+\frac{\gamma_1^2 k_0^4}{f_{-1}^2-f_0^2},  \label{varkappa}\\
\varkappa_0^2=f_0^2-\frac{\gamma_1^2 k_0^4}{f_{-1}^2-f_0^2},  \label{varkappa2}\\
f_n=\sqrt{\gamma_0k_0^2-\left(\beta+\frac{2\pi n}{a} \right)^2. \label{f}}
\end{eqnarray}
\begin{figure}[t]
\includegraphics[width=.44\textwidth, height=0.8\textwidth,trim={5.3cm 4.0cm 5.2cm 5.4cm},clip]{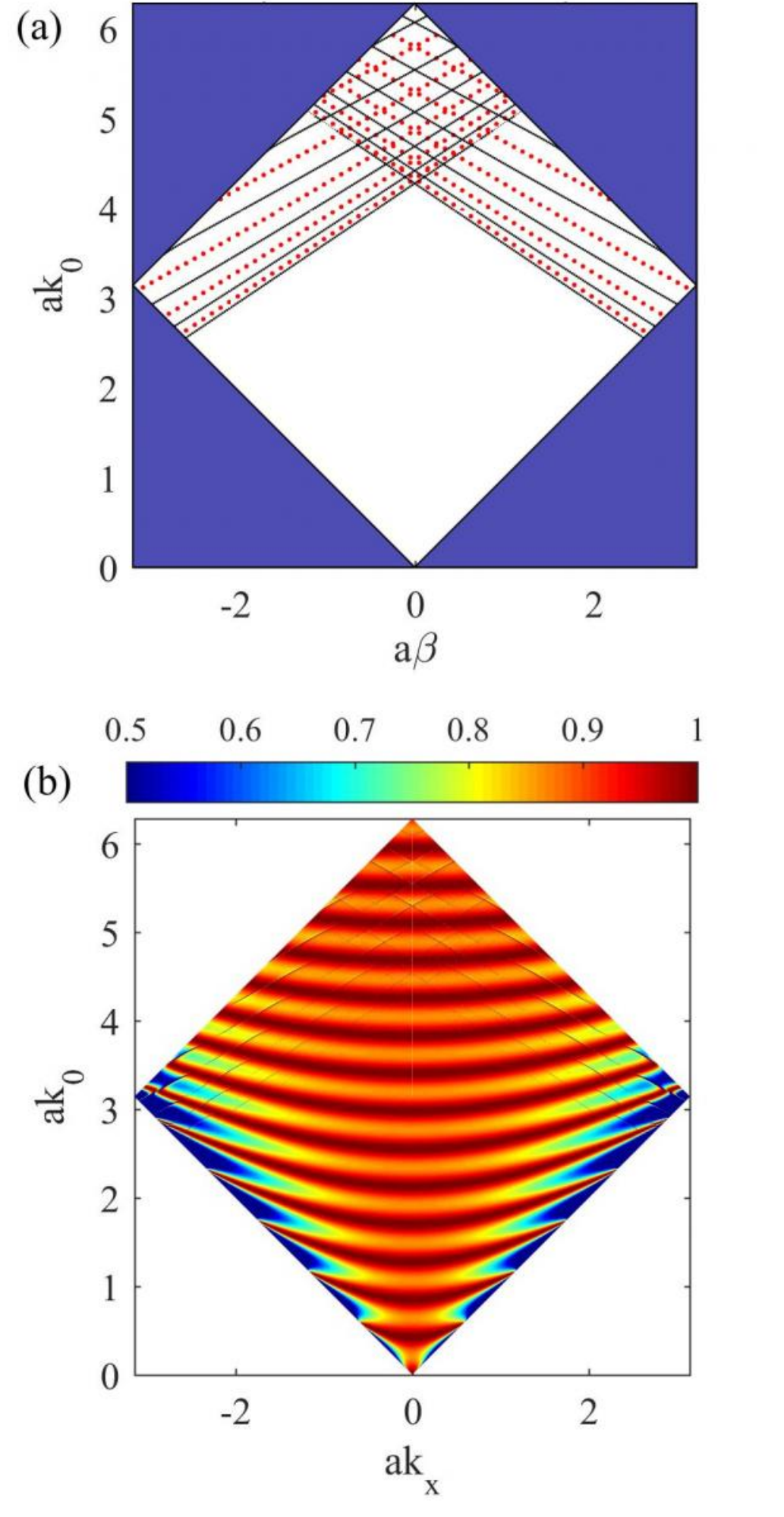}
\caption{(Color online) Spectra of low contrast DG, $h=5a, b=a/2, \epsilon_1=2.0736,\epsilon_2=2.2801$ (fused silica and soda lime glass at 1.861$\mu$m, respectively) (a)
Reduced bands on a dielectric slab with effective permittivity $\gamma_0$, solid
black - symmetric modes, red dots - antisymmetric modes. (b) Transmittance spectrum of the DG under illumination by a monochromatic
plane wave from the upper half-space obtained by RCWA.}\label{Fig2}
\end{figure}
A BIC is source-free solution decoupled from the open decay channel.
Setting $A=0,t_0=0$ in Eq. (\ref{big}) with non-zero $t_{-1}$ leads to
\begin{eqnarray}\label{condition1}
J_{0}=J_{-1},  \\\label{condition2}
-ik_{z,-1}=J_{-1}.
\end{eqnarray}
One the other hand, it immediately follows from Eqs. (\ref{condition1}, \ref{condition2}) that the determinant
of the matrix in Eq. (\ref{big}) is zero, and Eq. (\ref{big}) is solvable for $A=0$. Thus,
Eqs. (\ref{condition1}, \ref{condition2}) is the condition for BICs. By close examination of Eq. (\ref{big}) one
finds that in the limit $\gamma_1 = 0$ Eq. (\ref{condition2}) becomes
the dispersion equation of the reduced bands folded to the Brillouin zone for a uniform slab of permittivity $\gamma_0$, which according to Eq. (\ref{eps_Fourier})
is the average dielectric
permittivity of the DG.

We see now that in the two-wave approximation the BICs originate from the reduced guided modes on a uniform dielectric slab under an
 extra condition Eq. (\ref{condition1}). Essentially the same result could be obtained by considering a two mode approximation with
 account of $\gamma_{-1}$
 instead of $\gamma_{1}$, and $u_{1}$ instead of $u_{-1}$. This approximation yields the same formulas with $n=1$ instead of $n=-1$.
\begin{eqnarray}\label{condition3}
J_{0}=J_{1},  \\\label{condition4}
-ik_{z,1}=J_{1}.
\end{eqnarray}

\begin{figure}[t]
\includegraphics[width=.5\textwidth, height=.38\textwidth,trim={1.3cm 7.7cm 0.5cm 7.8cm},clip]{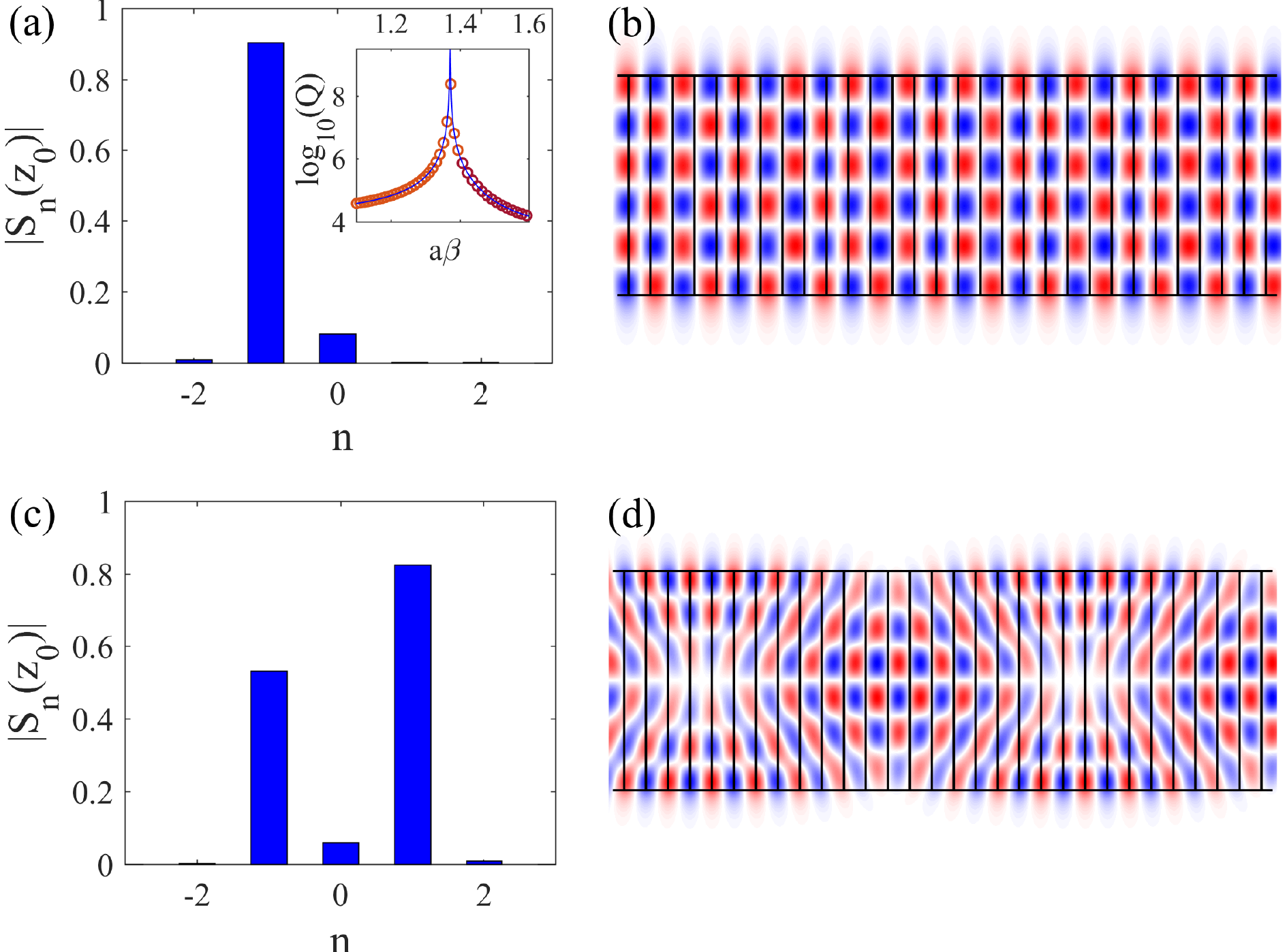}
\caption{(Color online) BICs in low-contrast DGs; two-wave BIC with found with RCWA $ak_0=4.0565, a\beta=1.3715$,  (a) bar diagram for $S_n(z_0)$, $z_0=h/6$; the inset shows
the dispersion of the $Q$-factor in the spectral vicinity of the BIC, solid blue - two-wave approximation, red circles - RCWA (b)
the BIC mode shape $Re\{\psi\}$; three-wave BIC with found with RCWA $ak_0=5.0941, a\beta=0.36712$, (c) bar diagram for $S_n(z_0)$, $z_0=h/6$ (d)
the BIC mode shape $Re\{\psi\}$.
The optogeometric parameters are the same as in Fig. \ref{Fig2}.}\label{Fig3}
\end{figure}
In Fig. \ref{Fig2} (a) we demonstrate the reduced bands (folded to the Brillouin zone)
of a uniform dielectric slab with the effective permittivity $\gamma_0$ in the
spectral range Eq. (\ref{range}). The bands are found by solving the dispersion equations (\ref{condition2}, \ref{condition4})
 with $\gamma_1=0$. The transmittance for the low-contrast DG under illumination by a plane wave from the upper
half-space computed with the use of RCWA is plotted in Fig. \ref{Fig2} (b).
By comparing Fig. \ref{Fig2} (a) against (b) one can see that each reduced guided mode on the dielectric slab corresponds
to an isolated high-quality resonance in the transmittance spectrum.
In the framework of RCWA the solution is found in the following form
\begin{equation}
\psi(x,z)=\sum_{n=-\infty}^{\infty}S_{n}(z)e^{ik_{x,n}x}.
\end{equation}
In \ref{Fig3} (a, b) we show the mode shape of a BIC hosted by a antisymmetric leaky
mode along with the bar diagram showing the expansion coefficients $S_{n}(z)$
obtained by RCWA.
One can see from Fig. (\ref{Fig3}) that $S_{0}(z),S_{-1}(z)$ dominate in the expansion. In the inset to Fig. (\ref{Fig3}) (a) we
demonstrate the dispersion of the $Q$-factor for the leaky mode hosting the BIC found with both RCWA and the two-wave approximation
Eq. (\ref{big}. Again, similarly to \cite{Gao17},
the two-wave mode predicts the behavior of the $Q$-factor to a good accuracy.

Finally, let us briefly discuss the accuracy of the approximation introduced in this section. It is seen from  Fig. \ref{Fig3} (a)
that other than $u_{-1}, u_0$ expansion coefficients, noticeably $u_{-2}$, are present in the RCWA solution. One can estimate
$u_{-2}$ by applying Rayleigh-Schr\"odinger perturbation theory \cite{Landau58a} to the zeroth order solution with
the only non-zero term $u_{-1}$. The result reads
\begin{equation} \label{u2}
u_{-2}=\frac{\gamma_1k_0^2}{f_{-2}^2-f_{-1}^2}u_{-1},
\end{equation}
while the same approach for $u_{0}$ yields
\begin{equation} \label{u0}
u_{0}=\frac{\gamma_1k_0^2}{f_{-1}^2-f_0^2}u_{-1}.
\end{equation}
The above equations predict $|u_{-2}/u_{0}|\approx 0.2$, and $|u_{-2}/u_{-1}|\approx 10^{-2}$ for the BIC shown in Fig. \ref{Fig3} (b). That is in
qualitative agreement with the data plotted in Fig. \ref{Fig3} (a). Numerically, the deviations are reflected
in a small difference between the position of the BIC found from RCWA ($ak_0=4.0565, a\beta=1.3715$), and
two-wave approximation Eqs. (\ref{condition1}, \ref{condition2}) ($ak_0=4.0576, a\beta=1.3708$).

\section{Three-wave BICs} \label{Section3}
The two-wave approximation breaks down in the spectral vicinity of the crossing between the guided modes in Fig. \ref{Fig2} (a).
In those points all three leading coefficients $\gamma_{-1}, \gamma_0, \gamma_1$ must be taken into account. It is technically
possible to find an analytical solution in such a three-wave approximation. This, however, results in awkward expressions
for the eigenvalues of $3 \times 3$ matrix. The analytical results are collected in the Appendix.
In this section we spare the reader of the cumbersome mathematics applying a generic scattering
theory for two spectrally close non-orthogonal resonances proposed by Volya and Zelevinsky \cite{Volya03}.
We mention in passing that the above approach is generic for two-mode settings \cite{SBR}.
For justification of applying  the formalism by Volya and Zelevinsky we again address the reader to
the Appendix.

The interference of two non-orthogonal resonances is described by the effective non-Hermitian operator
\begin{figure}[t]
\includegraphics[width=.5\textwidth, height=.48\textwidth,trim={2.1cm 6.6cm 3.cm 7.2cm},clip]{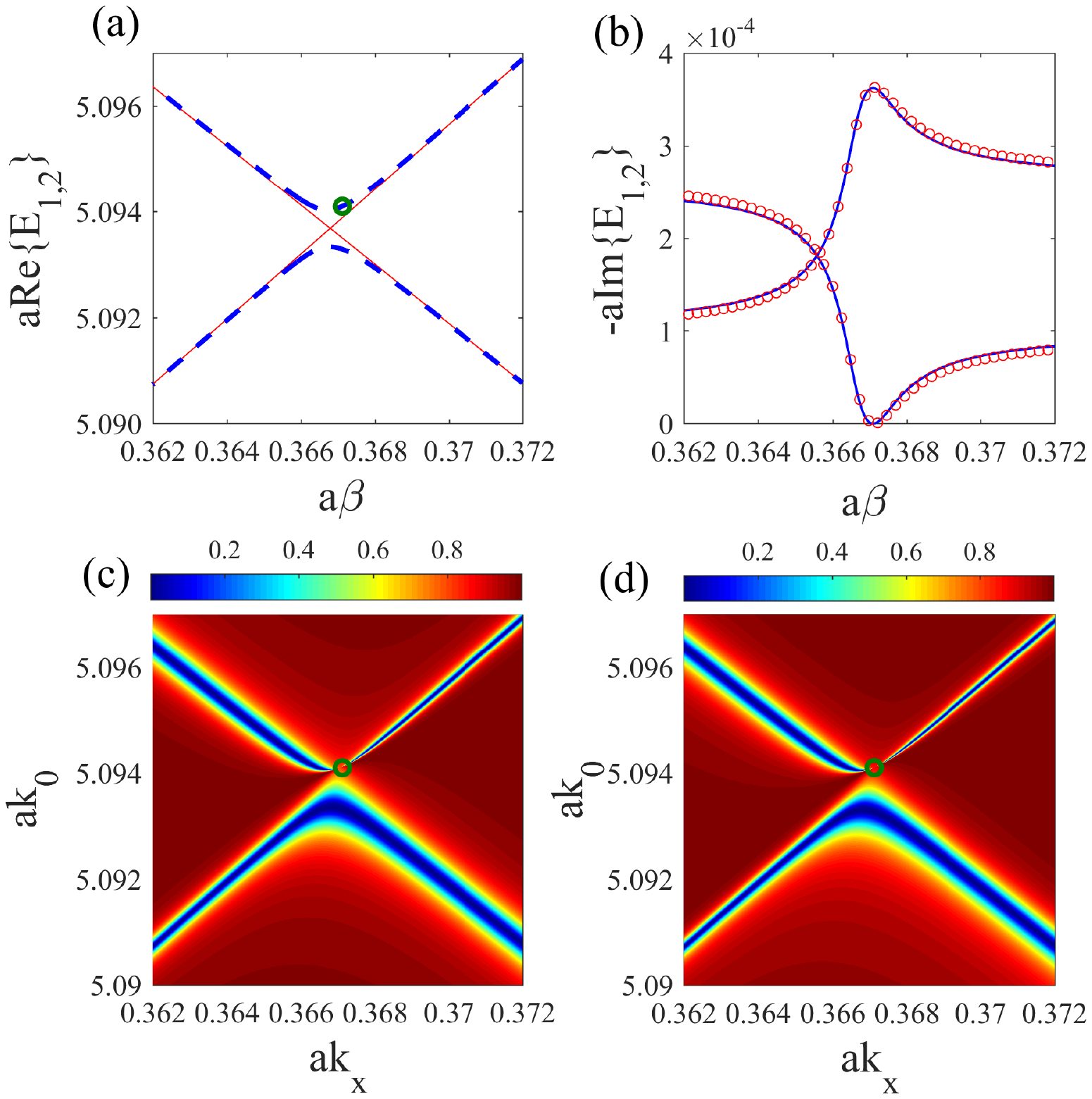}
\caption{(Color online) BIC in the spectral vicinity of an avoided crossing. (a) The real part of the eigenvalues of $\widehat{\cal H}$ with
$v=3.5 \times 10^{-4}$ - dash blue and the positions of the unperturbed resonances - solid red. (b)
the imaginary part of the eigenvalues of $\widehat{\cal H}$ with
$v=3.5 \times 10^{-4}$ - solid blue line, RCWA data - red circles. (c) The transmittance found from Eq. (\ref{trans_crossing}),
(d) The transmittance found with RCWA. The position of the BIC is shown by a green circle. The optogeometric parameters are the same as in Fig. \ref{Fig2}.}  \label{Fig4}
\end{figure}
\begin{equation}\label{Non-herm}
\widehat{\cal H}=
\left\{
\begin{array}{cc}
\omega_1 & v \\
v & \omega_{-1}
\end{array}
\right\}
-i\left\{
\begin{array}{cc}
\Gamma_{-1} & \sqrt{\Gamma_1\Gamma_{-1}} \\
\sqrt{\Gamma_{1}\Gamma_{-1}} & \Gamma_{-1}
\end{array}
\right\},
\end{equation}
where $\omega_{1,-1}$ are the real parts of the complex eigenfrequency of the unperturbed resonant leaky mode,
$\Gamma_{1,-1}$ are the corresponding resonant widths, and $v$ reflects the non-orthogonality between
the leaky modes. Notice that in the above equation we use the indices $1,-1$ rather than $1,2$ to underline the link with the expression presented in Appendix.
In the above definition the transmission amplitude is given by \cite{SBR}
\begin{equation}\label{trans_crossing}
t_{0}=2\frac{\omega(\Gamma_{1}+\Gamma_{-1})-\Gamma_{1}\omega_{-1}-\Gamma_{-1}\omega_{1}+v\sqrt{\Gamma_1\Gamma_{-1}}}{(\omega-E_1)(\omega-E_{2})},
\end{equation}
where $E_{1,2}$ are the eigenvalues of $\widehat{\cal H}$, while the condition for a BIC has the following form
\begin{equation}\label{position}
v(\Gamma_1-\Gamma_{-1})=\sqrt{\Gamma_1 \Gamma_{-1}}(\omega_1-\omega_{-1}).
\end{equation}

The positions of the unperturbed resonances $\omega_{1,-1}$ can be found from the transmittance spectra obtained by RCWA away of
the point of the avoided crossing. Then $\Gamma_{1,-1}$ can be found by solving Eq. (\ref{big}) in two-wave approximation outlined in the
previous section, or explicitly
from Eq. (\ref{width}). Thus, we are left with the only unknown fitting parameter $v$.
In Fig. \ref{Fig4} (a, b) we show the real and imaginary parts of the eigenvalues of $\widehat{\cal H}$ obtained by fitting Eq. (\ref{trans_crossing})
to the exact transmittance spectrum obtained by RCWA in the vicinity of an avoided crossing. The positions of the unperturbed resonances are also shown in Fig. (\ref{Fig4}) (a). The imaginary parts of the resonant eigenvalues obtained by RCWA are shown
in Fig. \ref{Fig4} (b). Finally, in Figs. \ref{Fig4} (c, d) we compare Eq. (\ref{trans_crossing}) against exact numerical
transmittance spectrum. One can see from Fig. \ref{Fig4} (c, d) that the results coincide to a good accuracy. The position of a BIC
given by Eq. (\ref{position}) also coincides with the numerical data. In Fig. \ref{Fig3} (c, d) we show the mode shape of the free-wave BIC along
with the bar diagram for $S_n(z)$. One can see that now three waves $n=-1,0,1$ dominate in the expansion.

\section{Conclusion}
In summary, we recovered the generic mechanism for BICs in low contrast DGs. It is shown
that the BICs originate from the reduced guided modes
on the effective dielectric slab with the permittivity equal
to the average permittivity of the DG. In case of isolated resonances the positions of BICs can be
found from two-wave dispersion relationships for guided leaky modes, as it is previously demonstrated
for fiber gratings \cite{Gao17}. In the case of the degeneracy between the two families of leaky modes the
system exhibits an avoided crossing of resonances.
We demonstrated that the occurrence of the avoided crossing can be quantitatively explained by a perturbative
solution of the tree-wave model. Numerically, in the spectral vicinity of the avoided crossing the transmittance as well
as the emergence of BICs
is well described in the framework of the generic formalism by Volya and Zelevinsky \cite{Volya03} with a single fitting
parameter. We speculate that ultra-high quality resonances previously reported in the literature \cite{Karagodsky11} can
be attributed to the emergence of BICs. Notice that Eq. (\ref{trans_crossing}) is always fulfilled in the $\Gamma$-point. Thus, the only condition for BIC
is the presence of leaky modes in the spectral range Eq. (\ref{range}). If the guided modes are supported on the slab with
the average permittivity $\gamma_0$, BICs always exist no matter how small the dielectric contrast is.
Recently, we have seen some interest to light localization and transmittance Fano
line shapes due to interference of two resonances \cite{Bykov15,Rybin17a,*Bogdanov18}. So far the link between avoided-crossings
and BICs has been mostly investigated in quantum systems \cite{Volya03,SBR,Pilipchuk17} and was only recently underlined
in the realm of optics \cite{Azzam18}.  We believe that the results presented may
be helpful in engineering BICs in optical set-ups, such as DGs.

This work was supported by Ministry of Education and Science of Russian Federation
(state contract N 3.1845.2017/4.6). We are grateful to K.N. Pichugin for assistance in
computations.

\bibliography{BSC_light_trapping}

\appendix*
\section{Analytical approach to three-wave BICs}

Let us consider analytical solution of Eq. (\ref{u_eqn}) when three Fourier components are taken into
account in Eq. (\ref{two_wave})
\begin{equation}
U_{s,a}(x)=u_{1}e^{i2\pi x/a}+u_0+u_{-1}e^{-i2\pi x/a}.
\end{equation}
In full analogue with Eq. (\ref{two_wave_eqn}) we have
\begin{eqnarray}\label{tree_wave_eqn}
f_0^2u_0+\gamma_1 k_0^2(u_{-1}+u_{1})=\varkappa^2 u_0, \nonumber \\
\gamma_1k_0^2u_0+f_{1}^2u_{1}=\varkappa^2 u_{1}, \nonumber \\
\gamma_1k_0^2u_0+f_{-1}^2u_{-1}=\varkappa^2 u_{-1},
\end{eqnarray}
where $f_{n}$ is given by Eq. (\ref{f}). Notice that the terms with $\gamma_{2}$ are absent Eq. (\ref{two_wave_eqn}) since $\gamma_2=0$
for $b=a/2$ as defined in Fig. \ref{Fig1}.
In the above equation we face an eigenvalue problem for $3\times 3$ matrix. Though
such a problem could be solved analytically with the use of Cardano's method that would result in cumbersome expressions
rendering the result unsuitable for further analysis. Here we restrict ourselves
with a perturbative analysis up to the terms ${\cal O} (\gamma_1^3)$. Then the tree solutions Eq. (\ref{tree_wave_eqn}) of the eigenvalue
problem read
\begin{eqnarray}
\varkappa_{-1}^2=f_{-1}^2+\frac{\gamma_1^2k_0^4}{f_{-1}^2-f_0^2}, \nonumber  \\
\varkappa_{1}^2=f_1^2+\frac{\gamma_1^2k_0^4}{f_{1}^2-f_0^2}, \nonumber  \\
\varkappa_{0}^2=f_0^2-\frac{\gamma_1^2k_0^4}{f_{-1}^2-f_0^2}-\frac{\gamma_1^2k_0^4}{f_{1}^2-f_0^2}. \label{tree_wave_solution}
\end{eqnarray}
The corresponding eigenvectors ${\bf u}_n= (u_{-1}, u_0, u_1)$ read
\begin{equation} \label{eigenvectors}
{\bf u}_{-1}=
\left(
\begin{array}{c}
1\\
\sigma_{-1}  \\
0
\end{array}
\right), \
{\bf u}_{1}=
\left(
\begin{array}{c}
0 \\
\sigma_{1}  \\
1
\end{array}
\right), \
{\bf u}_{0}=
\left(
\begin{array}{c}
\sigma_{-1}\\
-1 \\
\sigma_{1}
\end{array}
\right), \ \ \ \ \ \ \ \ \
\end{equation}
where
\begin{equation}\label{sigma_full}
\sigma_{n}=\frac{\gamma_1 k_0^2}{f_{n}^2-f_0^2}.
\end{equation}
Importantly, unlike Eq. (\ref{tree_wave_solution}) in Eq. (\ref{eigenvectors}) we only retained the terms up to ${\cal O} (\gamma_1^2)$.
This is because the second order Rayleigh-Schr\"odinger perturbation approach for eigenvectors results in bulky expressions \cite{Landau58a} which enormously
complicate further analysis.
Returning to Eq. (\ref{u_def}) one can write the general solution inside the slab as
\begin{eqnarray}\label{u_def_three_waves}
\psi(x,z)= e^{i\beta x}\left[ C_{-1}\left(\sigma_{-1}+e^{-i2\pi x/a} \right)F_{-1}(z) \right. \nonumber  \\
 +C_{1}\left(\sigma_{1}+e^{i2\pi x/a} \right)F_{1}(z) \nonumber  \\
\left.+C_0\left(1-\sigma_{-1}e^{-i2\pi x/a}-\sigma_{1}e^{i2\pi x/a}\right)F_{0}(z) \right],
\end{eqnarray}
where $C_{-1}, C_{0}, C_{1}$ are unknown coefficients to be defined from matching with the solution
outside the grating Eq. (\ref{outside}) and the functions $F_{n}(z)$ are given by
\begin{equation}
F_n(z)=\left\{
\begin{array}{l}
\frac{\cos(\varkappa_n z)}{\cos(\varkappa_n h/2)}, \ {\rm symmetric\ waves}\\
\frac{\sin(\varkappa_n z)}{\sin(\varkappa_n h/2)}. \ {\rm antisymmetric\ waves}
\end{array}
\right.
\end{equation}
Be applying the interface boundary conditions one can remove the unknowns $C_{-1}, C_{0}, C_{1}$ ending up
with a set of equation for the amplitudes of the outgoing waves $t_{-1} ,t_{0}, t_{1}$
\begin{widetext}
\begin{equation}\label{big2}
\left(
\begin{array}{ccc}
-J_0-J_{-1}\sigma_{-1}^2-J_{1}\sigma_{1}^2 & (J_{0}-J_{-1})\sigma_{-1} & (J_{0}-J_{1})\sigma_1\\
(J_{0}-J_{-1})\sigma_{-1} &-J_{0}\sigma_{-1}^2-J_{-1}(1+\sigma_{1}^2) &  (J_{-1}-J_{0})\sigma_{-1}\sigma_{1}\\
(J_{0}-J_{1})\sigma_1 & (J_{1}-J_{0})\sigma_{-1}\sigma_{1} &-J_{0}\sigma_{1}^2-J_{1}(1+\sigma_{-1}^2)
\end{array}
\right)
\left(
\begin{array}{c}
t_0+A \\
t_{-1} \\
t_{1}
\end{array}
\right)=i(1+\sigma_{-1}^2+\sigma_{1}^2)
\left(
\begin{array}{c}
k_{z,0}(t_0-A) \\
k_{z,-1}t_{-1} \\
k_{z,1}t_{1}
\end{array}
\right),
\end{equation}
where we again omitted all terms ${\cal O} (\gamma_1^3)$. The solution of Eq. (\ref{big2}) can be written in the following form
\begin{eqnarray}\label{tree_wave_final}
t_0=-A-\frac{2ik_{z,0}DA(\tilde{\Sigma}_{1}+\tilde{\Sigma}_{-1}+D\tilde{\Sigma}_{1}\tilde{\Sigma}_{-1})}
{Z_0}, \label{t0} \ \ \  \\
t_{-1}=-\frac{2ik_{z,0}DA}{\sigma_{-1}Z_{-1}}, \label{tm} \ \ \  \\
t_{1}=-\frac{2ik_{z,0}DA}{\sigma_{1}Z_{1}}, \label{tp} \ \ \
\end{eqnarray}
\begin{eqnarray}
Z_0=\tilde{E}_{1}(J_0-J_{-1}-\tilde{\Sigma}_0)+\tilde{\Sigma}_{-1}(J_0-J_{1}-\tilde{\Sigma}_0)-D\tilde{\Sigma}_0\tilde{\Sigma}_{1}\tilde{\Sigma}_{-1}, \label{Z0} \ \ \ \ \ \ \ \  \\
Z_{-1}=J_0-J_{-1}-\tilde{\Sigma}_0(1+\tilde{\Sigma}_{-1}D)+\frac{\tilde{\Sigma}_{-1}}{\tilde{\Sigma}_{1}}(J_0-J_1-\tilde{\Sigma}_0), \label{Zm} \ \ \ \ \ \ \ \  \\
Z_{1}=J_0-J_{1}-\tilde{\Sigma}_0(1+\tilde{\Sigma}_{1}D)+\frac{\tilde{\Sigma}_{1}}{\tilde{\Sigma}_{-1}}(J_0-J_{-1}-\tilde{\Sigma}_0) \label{Zp} \ \ \ \ \ \ \ \ \\
\tilde{\Sigma}_0=J_0+ik_{0,z}+J_{-1}\sigma_{-1}^2+J_{1}\sigma_{1}^2+ik_{0,z}(\sigma_{1}^2+\sigma_{-1}^2), \label{E0} \ \ \ \ \ \ \ \
\end{eqnarray}
\
\end{widetext}
and
\begin{eqnarray}
D=1+\sigma_{-1}^2+\sigma_{1}^2 \ \ \ \ \ \ \ \ \\
\tilde{\Sigma}_{-1}=\frac{\tilde{\epsilon}_{-1}}{\sigma_{-1}^2(J_0-J_{-1})} \label{Em}, \ \ \ \ \ \ \ \  \\
\tilde{\Sigma}_{1}=\frac{\tilde{\epsilon}_{1}}{\sigma_{1}^2(J_0-J_{1})}, \label{Ep} \ \ \ \ \ \ \ \ \\
\tilde{\epsilon}_n=ik_{z,n}+J_n.  \ \ \ \ \ \ \ \ \label{three_wave_def3}
\end{eqnarray}
Notice that after setting $\sigma_{1}=0$ Eq. (\ref{big2}) formally coincides
with Eq. (\ref{big}). On more rigorous grounds the two-wave approximation
can be justified by considering the quantities $\tilde{\Sigma}_{-1}, \tilde{\Sigma}_{1}$, which are generally diverging since $\sigma_{-1}, \sigma_{1}$
are vanishing with $\gamma_1$ according to Eq. (\ref{sigma_full}). This, however, is not the case if $\tilde{\epsilon}_{-1}\rightarrow 0$.
In that situation according to Eqs. (\ref{Zm}, \ref{Zp}) we have $|Z_{-1}|\gg |Z_{1}|$, and, consequently, $|t_{-1}|\gg |t_{1}|$
as one can see from Eq. (\ref{tm}, \ref{tp}). The latter inequality allows us to drop $t_{1}$ from Eq. (\ref{big2}). Remarkable,
the aforementioned condition, $\tilde{\epsilon}_{-1}\rightarrow 0$ is equivalent to Eq. (\ref{condition2}) as it is easily, seen from
Eq. (\ref{three_wave_def3}). The same arguments equally apply for the two-wave approximation with only $u_0, u_1$ taken into account,
when $\tilde{\epsilon}_{1}\rightarrow 0$. By extracting the imaginary parts of the denominator in Eqs. (\ref{tp}, \ref{tm}) one can find
the resonant widths in the spectral vicinity of the two-wave BICs in the following form
\begin{equation}\label{width}
\Gamma_n=\left.\frac{\sigma_{n}^2(J_0-J_n)^2 k_{z,0}}{|\tilde{\Sigma}_0|^2 \tilde{\epsilon}_n'}\right|_{k_0=\omega_0},
\end{equation}
where $\tilde{\epsilon}_n'$ is the derivative of $\tilde{\epsilon}_n$ with respect of $k_0$, and $\omega_0$ is the resonant
eigenfrequency.

\begin{figure}[ht]
\includegraphics[width=.4\textwidth, height=0.25\textwidth,trim={6.2cm 15.0cm 6.2cm 9.0cm},clip]{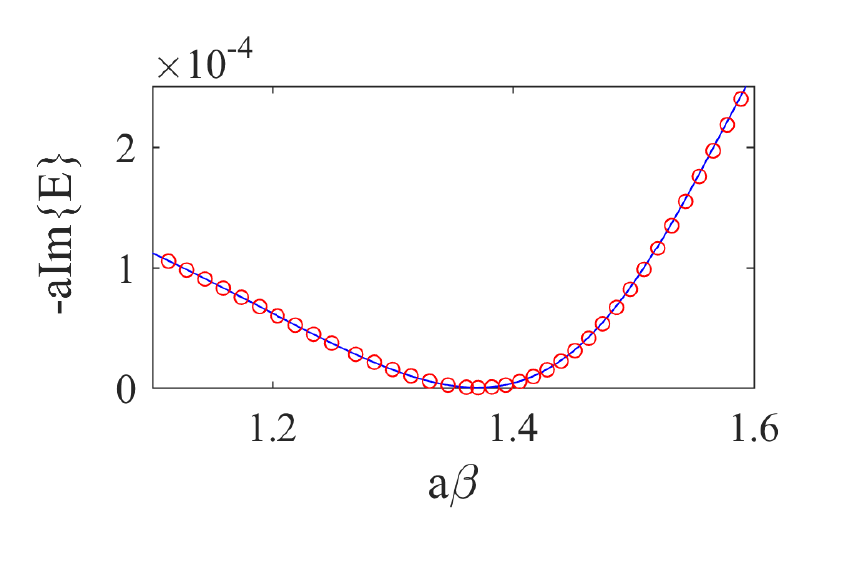}
\caption{(Color online) The imaginary part of the
resonant eigenfrequency for the leaky mode hosting the BIC from Fig. \ref{Fig3} (a), solid blue - RCWA data, red circles -
 Eq. (\ref{width}).}  \label{Fig5}
\end{figure}
In the spectral vicinity of a BIC the resonant properties of the DG are characterized the positions of the poles of the
reflection coefficient $t_0$, Eq. (\ref{t0}). The position of the poles $E_j$ correspond are the complex eigenvalues of Maxwell's equations
with real parts $Re\{E_j\}$ corresponding to the resonant frequency while $-Im\{E_j\}$ are the width of the resonances.
The poles can be found from analytic continuation of Eq. (\ref{Z0}) to the complex plane. In the case of isolated resonances
associated with two-wave BICs the analysis can be, however, simplified by allying Eq. (\ref{width}). In Fig. \ref{Fig5}
we plot the imaginary part of the resonant eigenvalue in the spectral vicinity of the BICs from
Fig. \ref{Fig3} (a) in comparison to the resonant width found from RCWA simulations. One can from Fig. \ref{Fig5} that
Eq. (\ref{width}) allows to find $\Gamma$ to a good accuracy since the position of the resonance $\omega_0$
is known from the dispersion equations (\ref{condition2}, \ref{condition4}) in the limit $\gamma_1\rightarrow 0$.

\begin{figure}[t]
\includegraphics[width=.4\textwidth, height=0.6\textwidth,trim={6.2cm 6.0cm 6.2cm 10.0cm},clip]{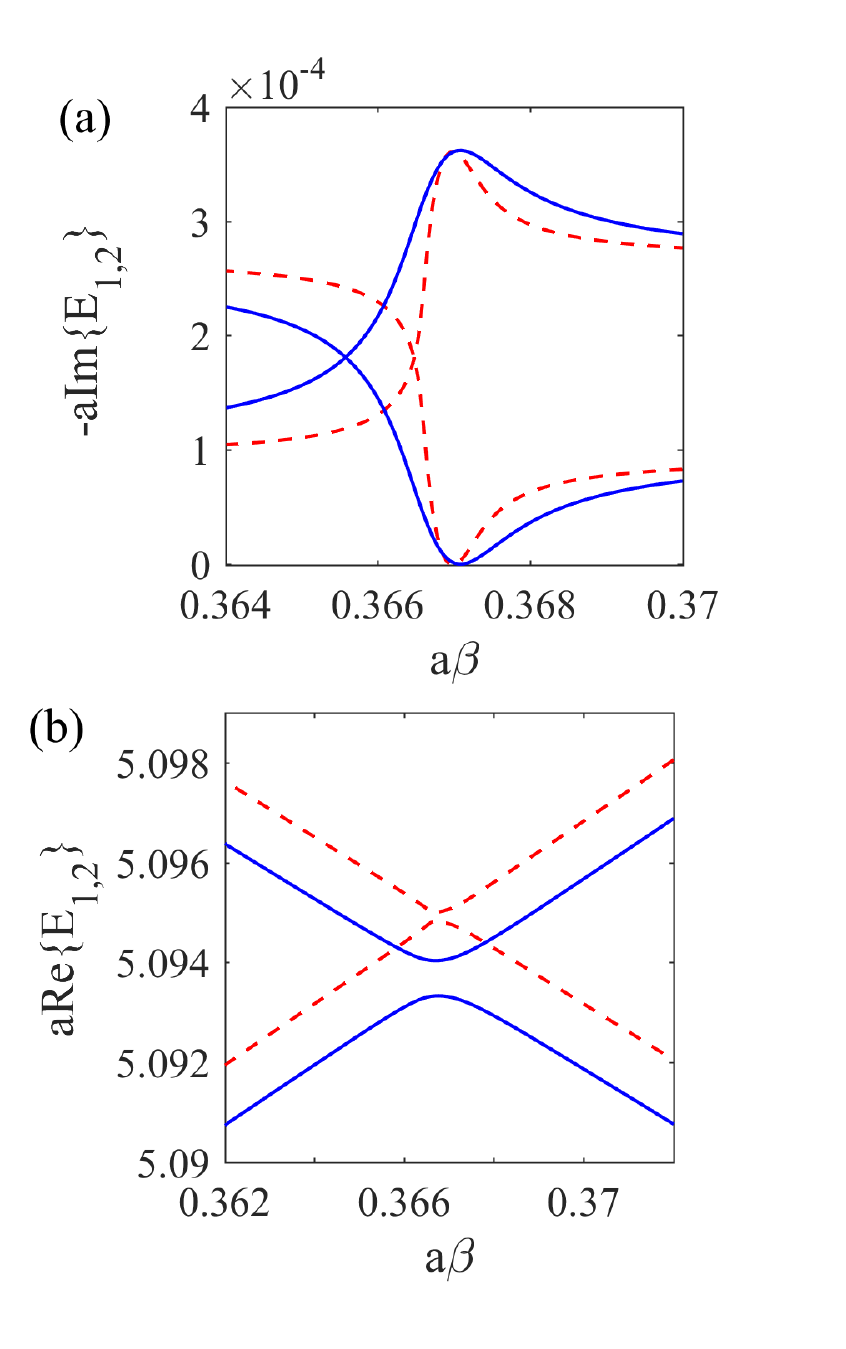}
\caption{(Color online) Resonant eigenvalues in the spectral vicinity of an avoided crossing. (a) The imaginary parts
of the resonant eigenvalues in the vicinity of the avoided crossing, solid blue - RCWA, dash red - analytic continuation
of Eq. (\ref{Z0}). (b) The real parts
of the resonant eigenvalues in the vicinity of the avoided crossing, solid blue - RCWA, dash red - analytic continuation
of Eq. (\ref{Z0}). }  \label{Fig6}
\end{figure}

The situation complicates, though, when both $\tilde{\Sigma}_{-1}, \tilde{\Sigma}_{1}$ become vanishing.
By recollecting
that Eqs. (\ref{condition2}, \ref{condition4}) with $\gamma_1 \rightarrow 0$ are the dispersion equations for the uniform dielectric slab
we immediately see that $\tilde{\epsilon}_n$, Eq. (\ref{three_wave_def3}) is small for $n=\pm 1$
at the intersection of the guided modes on the uniform dielectric slab. Then according to Eqs. (\ref{Em}, \ref{Ep})
both $\tilde{\Sigma}_{-1}, \tilde{\Sigma}_{1}$ are vanishing. In Fig. \ref{Fig6} we compare the
the resonant eigenvalues extracted from RCWA simulations against the position of the poles of the transmission coefficients,
Eq. (\ref{t0}). In both cases we found two eigenvalues $E_{1,2}$ with vanishing $\Gamma$. In Fig. \ref{Fig6} (a) we show the imaginary parts.
One can see that the two approaches are in qualitative agreement with one another, both predicting a vanishing resonant width.
The real parts of $E_{1,2}$ also demonstrate a qualitative agreement between the two approaches as seen from Fig \ref{Fig6} (b).
In fact, here we see an avoided-crossing between the position of the poles between typical for two leaky modes
interference mechanism of BICs proposed by Volya and Zelevinsky \cite{Volya03}.

Finally, a short remark is due on the accuracy of Eq. (\ref{t0}). By comparing Eq. (\ref{t0}) against Eq. (\ref{trans_crossing}) we find
the following expression for parameter $v$
\begin{equation}\label{v}
v=\frac{J_{1}+J_{-1}}{2k_{0,z}}+\frac{k_0}{J_0}.
\end{equation}
Numerically,  Eq. (\ref{v}) underestimates $v$ by approximately $3.5$ times in comparison with the exact spectrum. We speculate that
the deviations is due to ${\cal O} (\gamma_1^2)$ terms  dropped from Eq. (\ref{eigenvectors}). Nonetheless, the three-wave model analyzed
here is capable of both predicting the spectra of all isolated resonances to a good accuracy (see Fig. \ref{Fig5}), and qualitatively
describe their avoided crossing (see Fig. \ref{Fig6}).

\end{document}